\documentclass[%
 reprint,
 amsmath,amssymb,
 aps,
]{revtex4-2}

\usepackage[english]{babel}
\usepackage{graphicx}
\usepackage{xcolor}
\usepackage{dcolumn}
\usepackage{bm}
\usepackage{MnSymbol,wasysym}

\usepackage{multirow}

\begin{document}

\preprint{APS/123-QED}

\title{Quantum phase transitions of  tri-layer excitons in atomically thin heterostructures}

\author{Yevgeny Slobodkin$^{1}$, Yotam Mazuz-Harpaz$^{1}$, Sivan Refaely-Abramson$^{2}$, Snir Gazit$^{1,3}$, Hadar Steinberg$^{1}$ and Ronen Rapaport$^{1,*}$}
\affiliation{ 
\mbox{$^{1}$The Racah Institute of Physics, The Hebrew University of Jerusalem, Jerusalem 9190401, Israel}\\
\mbox{$^{2}$Department of Materials and Interfaces, Weizmann Institute of Science, Rehovot, Israel}\\
\mbox{$^{3}$The Fritz Haber Research Center for Molecular Dynamics,
The Hebrew University of Jerusalem, Jerusalem 9190401, Israel}\\
\mbox{$^{*}$Corresponding author: ronenr@phys.huji.ac.il}
}

\date{\today}

\begin{abstract}

We determine the zero temeperature phase diagram of excitons in the symmetric transition-metal dichalcogenide tri-layer  heterosctructure WSe$_{2}$/MoSe$_{2}$/WSe$_{2}$. First principle calculations reveal two distinct types of interlayer excitonic states, a lower energy symmetric quadrupole and a higher energy asymmetric dipole.
While interaction between quadrupolar excitons is always repulsive, anti-parallel dipolar excitons attract at large distances. 
We find quantum phase transitions between a repulsive quadrupole lattice phase and a staggered (anti-parallel) dipolar lattice phase, driven by the competition between the exciton-exciton interactions and the single exciton energies. 
Remarkably, the intrinsic nature of each interlayer exciton  is completely different in each phase.  This is a striking example for the possible rich quantum physics in a system where the single particle properties and the many-body state are dynamically coupled through the particle interactions.

\end{abstract}

\maketitle

\noindent\textbf{Introduction:} The quantum ground state of many-body systems is determined by a non-trivial interplay between inter-particle interactions and the delocalization induced by the kinetic energy. When interactions between particles are extended in space (e.g. in the case of dipolar particles), the many-body system may display significant particle correlations \cite{ astrakharchik_quantum_2007, buchler_strongly_2007,laikhtman_correlations_2009}. 
These many-body correlations can lead to exotic effects such as Roton instabilities \cite{chomaz_observation_2018} and super-solidity, both recently observed in ultra-cold gases of dipolar atoms  \cite{tanzi_observation_2019,chomaz_long-lived_2019,bottcher_transient_2019,guo_low-energy_2019,tanzi_supersolid_2019,natale_excitation_2019}. In semiconductor quasi-particle systems, collective correlated phases of interacting dipolar excitons in 2D heterostructures were observed, firstly in GaAs bilayers \cite{eisenstein_boseeinstein_2004,high_condensation_2012,shilo_particle_2013,stern_exciton_2014,anankine_quantized_2017,mazuz-harpaz_dynamical_2019}, and more recently in van der Waals (vdW) heterostructures ~\cite{wang_evidence_2019,sigl_condensation_2020}.

Quite generally, the form of interaction is encoded in the intrinsic properties of the elementary constituents determined by the internal structure, hence the interaction remains a static property. 
In typical realizations of dipolar atoms or excitons, the size and orientation of the dipole moment are fixed and set by an external field (magnetic or electric).
Specifically for spatially indirect (interlayer) excitons, which are quasi-particles formed by binding of excited electrons and holes residing in adjacent layers, the magnitude and orientation of the dipole-moment  are determined by the structural composition. This in turn dictates the exciton-exciton interactions. In this regard, heterostructure systems have an intrinsic decoupling between the single particle state and the collective state of many particles.

In this letter, we study excitons in a transition-metal dichalcogenide (TMD) trilayer heterostructure, and find a new phenomenon of  coupling between internal particle properties and the many-body state, making the single particle structure and the particle-particle interaction  dynamical parameters rather than static ones. 
We show that this unique situation gives rise to a quantum phase transition between two symmetry distinct phases, each made of completely different elementary exciton quasi-particles: a non-polar, weakly interacting many-body state at lower particle densities, a staggered dipolar state with strong interactions at higher densities, and a phase instability leading to a droplet phase.
\\

\begin{figure*}
\includegraphics[width=0.75\textwidth]{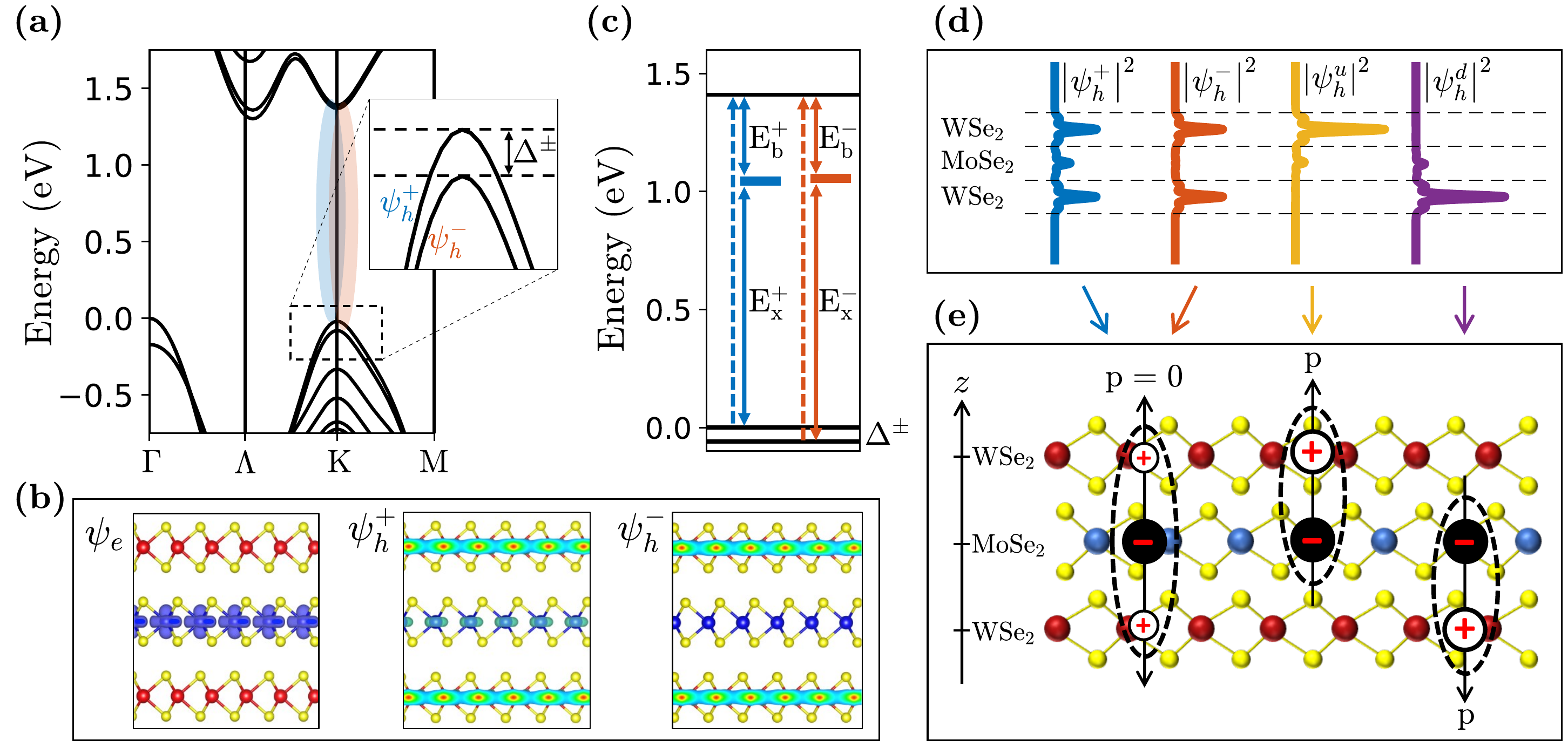}
\caption{\label{fig:GW-BSE} \textbf{(a)} Calculated GW quasiparticle bandstructure of the WSe$_2$/MoSe$_2$/WSe$_2$ heterostructure. The inset shows the valence energy split at the K region ($\Delta^\pm$).
\textbf{(b)} Quasiparticle wavefunctions of the two split valence bands at K, and the corresponding schematic representation of a double-well symmetric and anti-symmetric wavefunctions. \textbf{(c)} Exciton transition energies diagram for the two low lying excitons with energies $E_x^+$ and $E_x^-$, composed of $v\rightarrow c$ and $v-1\rightarrow c$ transitions, with binding energies $E_b^+$ and $E_b^-$, respectively. 
\textbf{(d)} Spatial cross-section along the $z$-direction of the probability density of the hole wavefunction  in the quadrupolar ($\left|\psi_h^{+\left(-\right)}\right|^2$) and dipolar exciton states  ($\left|\psi_h^{u\left(d\right)}\right|^2$).
\textbf{(e)} Schematic illustration of dipolar ($\text{p}\parallel z$ , $\text{p}\parallel -z$) and quadrupolar ($\text{p}=0$) exciton states. 
}
\end{figure*}

\noindent\textbf{Single quadrupolar and dipolar excitons:}
TMD heterostructures composed of stacked TMD monolayers have been shown to host long-lived excitons, a consequence of suppressed overlap between the wave-functions in separate layers~\cite{hong_ultrafast_2014, fang_strong_2014, rivera_observation_2015, Chiu_2015, rigosi_probing_2015, Unuchek_2018, rivera_valley-polarized_2016, merkl2019ultrafast}.
A commonly explored heterojunction is a bilayer composed of WSe$_2$ and MoSe$_2$ monolayers, where the conduction electrons and valence holes are localized at separate layers, resulting in interlayer excitons with a fixed, oriented electric dipole moment \cite{jin2018, jauregui2019, ciarrocchi2019, gillen2018, torun2018, ovesen2019interlayer, paik_interlayer_2019}.
Here we consider a trilayer stack, obtained by adding a second  WSe$_2$ layer - forming the WSe$_2$/MoSe$_2$/WSe$_2$ structure depicted in Fig.~\ref{fig:GW-BSE}(e). Trilayer stacks, studied in photoluminescence~\cite{baranowski_probing_2017,Choi_2018_Trilayer}, show indications of shortened exciton lifetime due to the enhanced wave-function overlap between the electron, which resides at the central MoSe$_2$ layer, and the hole, which is symmetrically delocalized between the two WSe$_2$ layers~\cite{Choi_2018_Trilayer}. The interplay between such a hole-delocalized, quadrupolar exciton and the bilayer dipolar excitons results in a rich many-body phase diagram.

To find the lowest lying exciton states, we start by computing the single particle band structure  using a many-body perturbation theory within the GW and Bethe-Salpeter equation (GW-BSE) approximation \cite{hybertsen1985first, rohlfing2000electron, deslippe2012berkeleygw}. The full computational details are given in \cite{SM}. Fig.~\ref{fig:GW-BSE}(a) shows the quasiparticle bandstructure of the trilayer system. 
Notably, the valence $K$ valley is split into upper and lower bands. The upper (with energy $E_h^+$) corresponds to the $z$-symmetric hole state $\psi_h^+$, which includes significant hole distribution on the MoSe$_2$ layer. The lower (with energy $E_h^-$) corresponds to the $z$-antisymmetric hole state $\psi_h^-$, where these contributions are absent (Fig.~\ref{fig:GW-BSE}(b)). 
The energy split between the two bands at the $K$-point (shown in the inset), $\Delta^{\pm}=E_h^+-E_h^-$, is analogous to that of the bonding and antibonding orbitals in a double-well system (see blue and red cross-sections in Fig.~\ref{fig:GW-BSE}(d)). 
As $\Delta^{\pm}$ depends on the interlayer distance $d$,  we allow for possible variation in $d$, yielding $\Delta^\pm = 20 - 60$ meV (see \cite{SM}). We note that the valence band split is not due to spin-orbit interactions, and exists in both spin channels.

An interlayer exciton is formed by e.g., a direct optical excitation of an electron from the valence to conduction band at the $K$-valley. We assume that the timescale for scattering into other valleys is long, and focus on $K$-valley processes alone. 
The lowest lying single exciton bound state, denoted $\psi_X^+$, with excitation energy $E_X^+=1.05$~eV (Fig.~\ref{fig:GW-BSE}(c)), is composed of a $K$-valley conduction electron state, $\psi_e$, spatially localized in the central MoSe$_2$ layer, and a $K$-valley valence hole state, $\psi_h^+$, with a wave-function delocalized in the $z$ direction between the symmetric top and bottom layers  (see \cite{SM}). The second exciton state, $\psi_X^-$, with excitation energy $E_X^-=1.1$~eV, is composed of $\psi_e$ and $\psi_h^-$. 
Contrary to the bilayer exciton picture, $\psi_X^{\pm}$ have a zero dipole and a finite quadrupole moment (shown schematically in Fig.~\ref{fig:GW-BSE}(e), left) and are therefore named quadrupolar excitons. 
The exciton binding energies of the two quadrupolar states are given by $E_b^\pm = E_e+E_h^\pm-E_X^\pm $.   
Our calculation shows that the binding energies $E_b^+$ and $E_b^-$, for the symmetric and antisymmetric quadrupole excitons, respectively, are almost identical ($0.36$~eV and $0.35$~eV). The difference between the energies of two quadrupolar excitons is therefore very close to the valence band split, $E_X^+-E_X^- \approx \Delta^\pm$.

We now turn to the  construction of the $K$-valley dipolar states, $\psi_X^u$ and $\psi_X^d$,  as in a double-well model, as shown in Fig.~\ref{fig:GW-BSE}(d).
To do that, we construct  hole states localized in the top (bottom) layer, $\psi_{h}^{u(d)}$, by adding and subtracting the two delocalized hole states: \(\left|\psi_h^{u(d)}\right\rangle=\frac{1}{\sqrt{2}}\left(\left|\psi_h^+\right\rangle\pm\left|\psi_h^-\right\rangle \right)\) (where we assume opposite phases at the different WSe$_2$ layers in $\psi_{h}^-$), with energies $E_{h}^u=E_{h}^d=\frac{1}{2}\left(E_h^+ + E_h^-\right)$. 
The dipolar exciton states $\psi_X^{u(d)}$ are then constructed by an electron state $\psi_e$ and a hole state $\psi_h^{u(d)}$, localized in the upper (lower) layer: $\left|\psi_X^{u(d)}\right\rangle =\frac{1}{\sqrt{2}}\left(\left|\psi_X^+\right\rangle \pm\left|\psi_X^-\right\rangle \right)$. These excitons, depicted on the right side of Fig.~\ref{fig:GW-BSE}(e) have a non-vanishing electric dipole moment, each pointing in an opposite direction. 
Since $E_b^{+}\approx E_b^{-}$, the binding energies of the two types of dipolar excitons, $E_b^{u(d)}$, are approximately equal to those of the quadrupolar excitons. 
Thus the energy of both dipolar exciton states is given by $E_X^{u(d)} = E_e+E_h^{u(d)} -E_b^{u(d)}\approx E_e+E_h^{u(d)} -E_b^{\pm}$. 
The energy gap between the hole-symmetric quadrupolar  exciton and the dipolar excitons, $\Delta_{DQ}$, can therefore be evaluated to be:  
\begin{equation}
    \Delta_{DQ}=E^{u(d)}_X-E^{+}_X
    \approx E_h^{u(d)}-E_{h,+} =\Delta^\pm/2, 
\end{equation}
which yields $\Delta_{DQ}=10-30$meV for the above layered structure. 
In what follows, we will show that $\Delta_{DQ}$ is an important parameter  affecting the possible ground states in the many-exciton limit. 
Since $E_X^{+}<E_X^{u(d)}<E_X^{-}$, in the following we only focus on the lowest energy hole-symmetric quadrupole exciton $\psi_X^+$, and the two degenerate dipolar excitons, $\psi_X^{u(d)}$.\\

\noindent\textbf{Finite density of excitons: phase transitions at T=0:}
With the quadrupolar and dipolar single exciton states established, we now turn to discuss the many-body phase diagram at finite exciton density, $n$.  

In constructing our effective low-energy description, we consider only the dilute exciton limit, where the typical inter-exciton distance is significantly larger than the exciton size (set by the electron-hole bound state).
This allows us to treat excitons as point-like bosonic quasi-particles and safely neglect corrections arising due to fermionic exchange \cite{mazuz-harpaz_dynamical_2019}.
The two dipolar states $\psi_X^{u,d}$ can be conveniently parametrized by an Ising degree of freedom $|\sigma^z=u,d\rangle$. The exciton dynamics is governed by the Hamiltonian (setting $\hbar=1$),
\begin{equation}
    \mathcal{H}=-\frac{1}{2m_x}\sum_{i} \nabla_i^2+\sum_{\substack{i<j}}V^{d}_{\sigma^z_i,\sigma^z_j}(|r_j-r_i|)-\Delta_{QD}\sum_i \sigma_i^x.
    \label{eq:MB_h}
\end{equation}

In the above equation, $i$ labels excitons and $m_x$ denotes the effective in-plane exciton mass. The layer dependent dipolar interaction reads:
\begin{equation}
\begin{aligned}
V^d_{\sigma_i^z,\sigma^z_j}(r)\;\;&=V_{p,p}(r)\delta_{\sigma_i^z,\sigma_j^z}+V_{p,-p}(r)\delta_{\sigma_i^z,-\sigma_j^z},\\
V_{p,p}\left(r\right)\;\;&=\frac{e^{2}}{\kappa}\left(\frac{2}{r}-\frac{2}{\sqrt{r^{2}+d^{2}}}\right)\\
V_{p,-p}\left(r\right)&=\frac{e^{2}}{\kappa}\left(\frac{1}{r}+\frac{1}{\sqrt{r^{2}+\left(2d\right)^{2}}}-\frac{2}{\sqrt{r^{2}+d^{2}}}\right)
\end{aligned}
\end{equation}
Here, $V_{p,p}$ and $V_{p,-p}$ denotes the electrostatic energy associated with a parallel ($(u,u)$ or $(d,d)$) and anti-parallel ($(u,d)$ or $(d,u)$) configuration of two dipolar excitons. $d$ is the inter-layer distance and $\kappa$ is the effective dielectric constant, taken from our GW calculation as the dielectric function at the interaction distance corresponding to the interlayer separation \cite{SM}).
The operator $\sigma^x|u\rangle=|d\rangle$, $\sigma^x|d\rangle=|u\rangle$  locally flips the dipole moment orientation. We note that the Hamiltonian affords a 
$\mathbb{Z}_2$ Ising symmetry corresponding to a global flip of the dipole moment orientation $\sigma^z_i\to-\sigma^z_i$ for all $i$.

To highlight the interplay between dipolar and quadrupolar states, we quench the in-plane exciton dynamics, namely $e^2\sqrt{n}/\kappa m_x\ll1$. In this limit, the ground state is determined by a competition between the electrostatic energy and the quantum  dynamics of the dipole orientation, as described by the second and third terms in Eq.~\eqref{eq:MB_h}, respectively. 

The typical scale of the potential energy term of the Hamiltonian is $e^2/\kappa d$, and of the quantum dynamics is $\Delta_{DQ}$, suggesting a dimensionless parameter as their ratio: $R=\Delta_{DQ}\kappa d/e^2$. The zero-temperature ground state of the many particle system is determined in a two-parameter phase diagram controlled by $R$ and $n$. To gain insight into the possible competing phases, we begin by establishing the ground state configuration in the various parameter limits, and then analyse the boundaries between these phases. 

\begin{figure}
\includegraphics[width=0.48\textwidth]{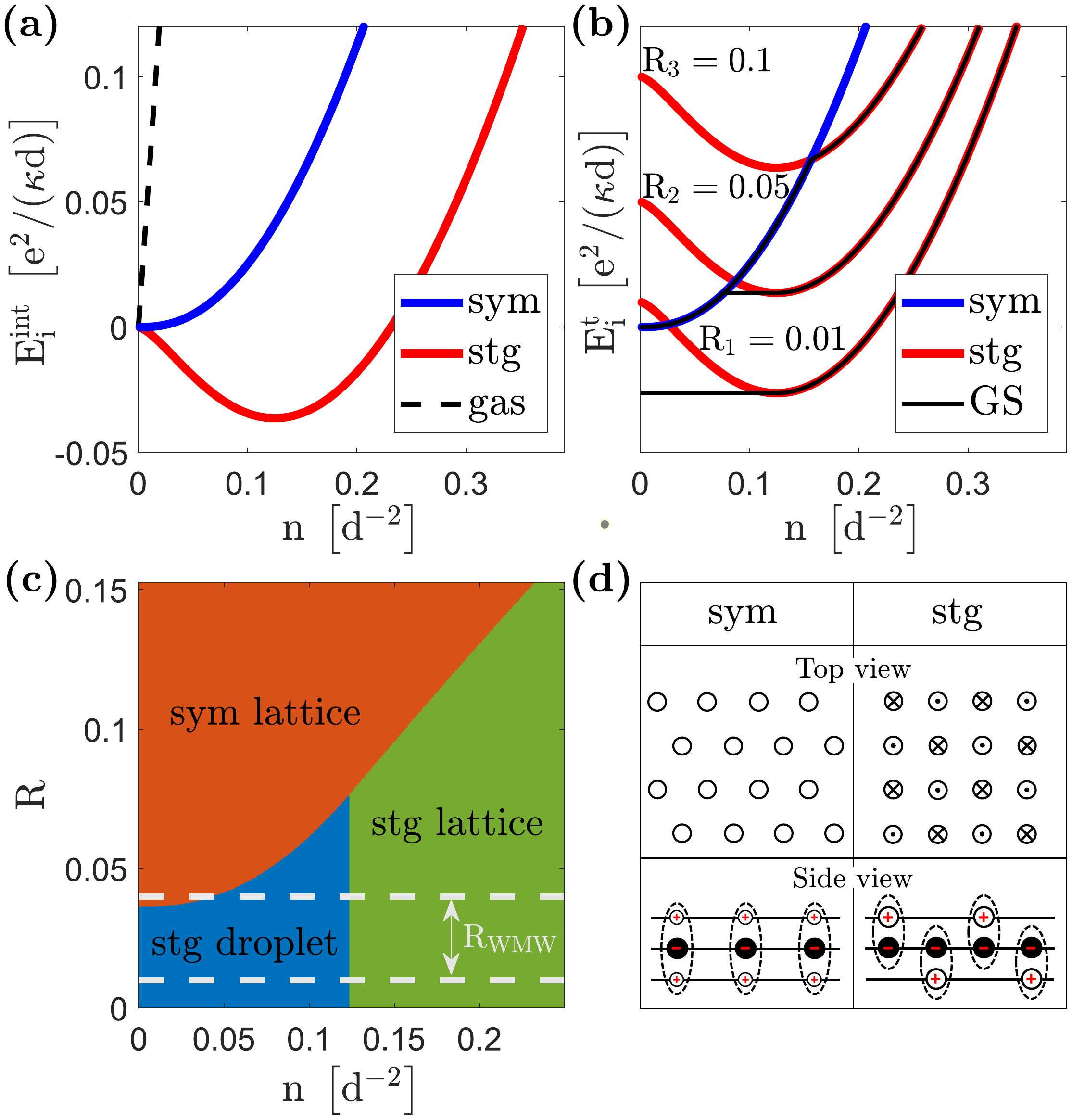}
\caption{\label{fig:Phase_diagram} 
\textbf{(a)} Electrostatic interaction energy $E_i^{\text{int}}$ vs. the density of excitons $n$. stg represents a staggered square lattice of dipolar excitons, where nearest neighbors have anti-parallel dipole moments. sym represents a triangular lattice of quadrupolar excitons. The minimum energy (stg) appears at $n_{d}\approx0.12d^{-2}$. Dashed black represents an uncorrelated exciton gas (see \cite{SM}).
\textbf{(b)} Total energy per particle $E_i^t$ (where $i=\text{sym/stg}$) vs. the density of excitons $n$. GS represents the many body ground state (black lines). Three different phase transitions are displayed: stg droplet$\rightarrow$stg lattice ($R_1=0.01$), sym lattice$\rightarrow$stg droplet$\rightarrow$stg lattice ($R_2=0.05$), sym lattice$\rightarrow$stg lattice ($R_3=0.1$). \textbf{(c)} Phase diagram. $R$ represents the ratio of the energy difference between a dipolar and a quadrupolar single exciton state and the potential energy scale $e^2/\kappa d$. Dashed gray lines represent the upper and lower estimates for $R_{WMW}$, the special case of a WSe$_2$/MoSe$_2$/WSe$_2$ tri-layer. \textbf{(d)} Schematic illustration of the competing phases in a three-layer structure with dipolar and quadrupolar excitons. }
\end{figure}

First, we consider the limiting case $R\to\infty$ for any finite density $n$. In this case the last term in Eq.~\eqref{eq:MB_h} dominates. Consequently, strong quantum fluctuations in the dipole moment orientation favor a fully quadrupolar ground state, $\left|\psi_X^{+}\right\rangle =\left|+\right\rangle=\frac{1}{\sqrt{2}}\left(\left|u\right\rangle +\left|d\right\rangle\right)$. Since quadropolar  electrostatic interactions, $(+)\leftrightarrow (+)$, are purely \emph{repulsive}, minimizing the electrostatic energy, leads to a triangular lattice structure \cite{SM}  with broken translational symmetry, see left panels of Fig.~\ref{fig:Phase_diagram}d. In Fig.~\ref{fig:Phase_diagram}a we plot the total electrostatic energy per exciton, $E^{\text{int}}_{\text{sym}}(n)$ in the above configuration. It is simple to show that it rises as $E^{\text{int}}_{\text{sym}}\sim n^{5/2}$ at low density, $nd^2\ll 1$ (\cite{SM}).
	
Next, we examine the high density $nd^2\to\infty$ limit for finite $R$, for which the electrosatic energy contribution in Eq.~\eqref{eq:MB_h} overwhelms all other terms and hence should be minimized. A key observation is that in contrast to quadrupoles which always repel, anti-parallel dipoles, $(u)\leftrightarrow (d)$, \emph{attract} at long distances leading to an electrostatic energy gain at all densities compared to the symmetric case, as is seen in Fig.~\ref{fig:Phase_diagram}a. 
Therefore at this limit, the system favours a competing dipolar configuration comprising a staggered pattern of dipole orientations, see the right panels of Fig.~\ref{fig:Phase_diagram}d. The particular choice of a square lattice (as opposed to a triangular lattice) allows avoiding the inherent frustration of staggered configurations on non-bipartite lattices, \cite{sammon_attraction_2019}. 
Importantly, the staggered state not only breaks translational symmetry, but also the \emph{Ising layer symmetry}.
The total electrostatic energy per exciton of the staggered state, $E^{\text{int}}_{\text{stg}}(n)$, evolves non-trivially as a function of density, see Fig.  \ref{fig:Phase_diagram}a. At low densities, the energy turns negative,     $E^{\text{int}}_{\text{stg}}(n)\sim-n^{3/2}$. However, with increasing $n$, the interaction energy increases and turns positive, due to the short range repulsive interaction. The energy minimum is obtained at $n_d\approx0.12d^{-2}$, at which the interaction energy equals $E^{\text{int}}_{\text{stg}}(n_d)\approx-0.036e^2/(\kappa d)$.

Next, we turn to determine the phase boundaries separating the staggered dipolar and layer-symmetric quadrupolar states. To that end, we examine the total energy per exciton $E^i = E_X + E^{int}(n)$, and compare the total energies in the two competing states:  $E^t_{sym}$ vs. $E^t_{stg}$. Clearly, this variational approach is approximate, as it neglects quantum fluctuations. We leave the question of determining their role to a future study \cite{Eran_unpublished}. 

In the layer-symmetric state the energy is simply given by $E^t_{\text{sym}}(n)=E^{\text{int}}_{\text{sym}}(n)$, where for convenience we set the overall energy reference scale, $E_X^+=0$, to zero. By contrast, in the staggered dipolar lattice, in addition to the electrostatic energy, one must also take into account the energy contribution of the quadropolar to dipolar gap $\Delta_{DQ}$, so that in total $E^t_{\text{stg}}(n)=\Delta_{DQ}+E^{\text{int}}_{\text{stg}}(n)$. 
This energy minimization procedure is illustrated in Fig.~\ref{fig:Phase_diagram}b, where we depict both $E^t_{\text{sym}}$ and $E^t_{\text{stg}}$ as function of $n$ for several pertinent values of $R$.

We first consider the case of large but finite $R$. Since  $\Delta_{DQ}$ and thus $R$ are strictly positive, for sufficiently large $R$ and at low $n$, where interactions between excitons are negligible, the symmetric quadrupole state $\psi_X^+$ will always be the lowest energy state, as already discussed above. With an increase of $n$, the electrostatic energy gain associated with the staggered configuration eventually overwhelms $\Delta_{DQ}$. Therefore, we expect to find a quantum phase transition, at a critical density $n_c(R)$, where the symmetric quadropolar state gives way to a staggered dipole configuration. 
Neglecting quantum fluctuations, the phase transition is expected to be first order in nature (as opposed to the pure Ising universality class), since in addition to breaking of Ising layer-symmetry the transition also involves a structural rearrangement, from a triangular to square lattice. The precise phase boundary $n_{c}(R)$ are determined by carrying out a numerical computation \cite{SM}.

The above picture is correct for all $R$-values for which $n_c>n_d$. Interestingly, we further identify an additional phase for $R<R_c=R(n_c=n_d)\approx0.076$. This result follows directly from the non-monotonous behavior of $E_{\text{stg}}^{\text{int}}(n)$. As is depicted in Fig.~\ref{fig:Phase_diagram}b, for any $R<R_c$, there is a density $n_c<n_d$ above which  $E_{\text{stg}}(n_d)<E_{\text{sym}}(n)$. This suggests that a homogeneous symmetric state with a density $n>n_{c}$ is unstable towards a phase separation and formation of a staggered dipolar droplet with a density $n_d$.
The phase separation consisting of a staggered dipolar droplet is sustained up to $n=n_d$, beyond which the droplet fills the plane and the homogeneous staggered phase is reached. 
Surprisingly, for $R<0.036$, $n_c=0$, and the symmetric quadrupole phase is unstable for any density. Schematic drawing and further analysis of the staggered droplet phase is given in \cite{SM}.

Fig.~\ref{fig:Phase_diagram}c presents the general phase diagram of a three-layer system as a function of $R$ and $n$. Within our approximation, the triple point appears at $(n_d,R_c)$. Dashed lines represent the predicted possible range of $R_{WMW}$ and thus the possible phases, for the special case of a system of WSe$_2$/MoSe$_2$/WSe$_2$ tri-layer calculated above. This range results from the uncertainty in calculating both $\Delta_{WMW}$ and $\kappa$, as discussed above \cite{SM}. This clearly demonstrates that such quantum phase transitions are indeed relevant for the TMD trilayer system. \\

\noindent\textbf{Discussion and summary:} A complete understanding of the low temperature phase diagram of our model will require a refined analysis that takes into account the role of quantum fluctuations and bosonic exchange statistics. These effects may allow access to additional states of matter such as various patterns of exciton condensates and super-solids. Importantly, our many-body Hamiltonian, Eq.~\eqref{eq:MB_h}, is amenable to an exact numerical solution using quantum Monte Carlo techniques. These interesting directions are currently under pursuit \cite{Eran_unpublished}.
 
 From the experimental perspective, the symmetric and staggered  phases can be identified in optical spectroscopy through the difference in radiative rates of quadrupolar and dipolar excitons \cite{Choi_2018_Trilayer}, as well as through the different scaling of the exciton recombination energy with density. The droplet phase would be characterized by a regime where the recombinaton energy is density independent. Our theoretical predictions (fig.~\ref{fig:Phase_diagram}c) are not unique to the specific WSe$_{2}$/MoSe$_{2}$/WSe$_{2}$ heterostructure. In principle, it can be realized in many other tri-layer heterostructures of two materials that form a type-II band alignment, and also in three monolayers of the same material separated by insulating spacer layers, with a bias applied between the middle layer and the two lateral layers. Similar potentials can also be designed in semiconductor quantum wells based on, e.g., GaAs and AlAs compounds. In addition, properties such as the dipole length and $\Delta^{\pm}$, are tunable in each specific structure (by applying an electric gate or pressure). Experiments with different three-layer systems may explore different regimes of the phase diagram.

To summarize, we present a tri-layer system as a striking example for the possible rich quantum physics in a system where the single particle properties and the many-body state are no longer separate entities but rather dynamically coupled through the particle interactions. In the single exciton limit we predict the emergence of a new type of interlayer excitons with a finite electric quadrupole moment. At finite exciton densities, our simple model suggests unique phase transitions that change both the Ising (layer) and lattice symmetries, and the intrinsic nature of each interlayer exciton.

\begin{acknowledgments}
SRA acknowledges support from the Israel Science Foundation, Grant No.1208/19. SG acknowledges support from the Israel Science Foundation, Grant No. 1686/18. RR acknowledges support from the Israel Science Foundation, Grant No. 836/17, and the Bi-national Science Foundation, Grant No. 2016112. HS acknowledges support from Israeli Science Foundation grant 861/19. This research used resources of the National Energy Research Scientific Computing Center (NERSC).
\end{acknowledgments}
\newpage

\bibliography{main}

\end{document}


\preprint{APS/123-QED}

\title{Supplemental materials:\\Quantum phase transitions of  tri-layer excitons in atomically thin heterostructures}

\author{Yevgeny Slobodkin$^{1}$, Yotam Mazuz-Harpaz$^{1}$, Sivan Refaely-Abramson$^{2}$, Snir Gazit$^{1,3}$, Hadar Steinberg$^{1}$ and Ronen Rapaport$^{1,*}$}
\affiliation{ 
\mbox{$^{1}$The Racah Institute of Physics, The Hebrew University of Jerusalem, Jerusalem 9190401, Israel}\\
\mbox{$^{2}$Department of Materials and Interfaces, Weizmann Institute of Science, Rehovot, Israel}\\
\mbox{$^{3}$The Fritz Haber Research Center for Molecular Dynamics,
The Hebrew University of Jerusalem, Jerusalem 9190401, Israel}\\
\mbox{$^{*}$Corresponding author: ronenr@phys.huji.ac.il}
}

\date{\today}

\keywords{Excitons, Transition Metal Dichalcogenides, van der Waals heterostructures, Phase transition, Many-body physics}

\maketitle

\section{\label{Sec:GW-BSE}GW-BSE computational details}

To obtain the starting point for our GW calculation, we first perform a density functional theory (DFT) calculation in the local density approximation (LDA) \cite{Kohn1965}, with the Quantum Espresso package \cite{QE-2017}. The calculations were done with a periodic unit cell, in which all TMD monolayers are assumed to have the same lattice vector and within a AA' stacking arrangement, as shown in Fig.~\ref{fig:gwbse}. We use a plane-wave basis and norm-conserving pseudopotentials. The distance between repeated unit cells in the out-of-plane direction was 25$\AA$. We fully relaxed the atomic coordinates while constraining the lattice constant to the experimental values of 3.29$\AA$  for both WSe$_2$ and MoSe$_2$ monolayers. A $30\times30\times1$ k-point grid was used to calculate the self-consistent charge density  with a 70 Ry wave function cutoff.  We included spin polarization at the DFT level using fully-relativistic corrections. 

To set the interlayer distance, we used three different DFT approximations- pure LDA; LDA plus vdW corrections within the van der Waals density functional with Cooper exchange (vdW-DF-c09x) \cite{cooper2010van, Dion2004}, and PBE \cite{PBE1996} plus Grimme DFT-D3 van der Waals corrections scheme \cite{grimme2010}. These approximations results in W-M interlayer distances of 6.13$\AA$, 6.20$\AA$, and 6.60$\AA$, respectively, and influence the valence band splitting at K, as we discuss in the main text. The valence energy-split value ranges between 20, 52, and 60 meV for the interlayer separations computed with the DFT-D3 vdW functional, the  vdW-DF-c09x functional, and the non-corrected LDA funcitonal.

We computed the quasiparticle bandstructure using the GW method within the BerkeleyGW code \cite{deslippe2012berkeleygw}. Our results are obtained within the Hybertsen-Louie generalized plasmon-pole (HL-GPP) model \cite{deslippe2012berkeleygw, hybertsen1985first}. In the GW calculation, we used an energy cutoff of 25 Ry for the reciprocal lattice components of the dielectric matrix and included 5000 bands in the summation. We employed the nonuniform neck subsampling (NNS) scheme \cite{felipe2017nonuniform} to sample the Brillouin zone and to speed up the convergence with respect to k-point sampling. In this scheme, we use a $6\times6\times1$ uniform q-grid and include an additional set of 10 q-points in the Voronoi cell around q=0. This corresponds to an effective q-grid of more than 1000x1000x1 uniform q-points and converges the QP energies within $\sim$0.01 eV. A truncated Coulomb interaction was used to prevent spurious interactions between periodic images of the 2D sheet \cite{ismail2006truncation}. Spin orbit coupling is included as a perturbation following Refs.~\cite{Qiu2013,Qiu2016}.

\renewcommand{\thefigure}{S1}
\begin{figure}
\includegraphics[width=0.45\textwidth]{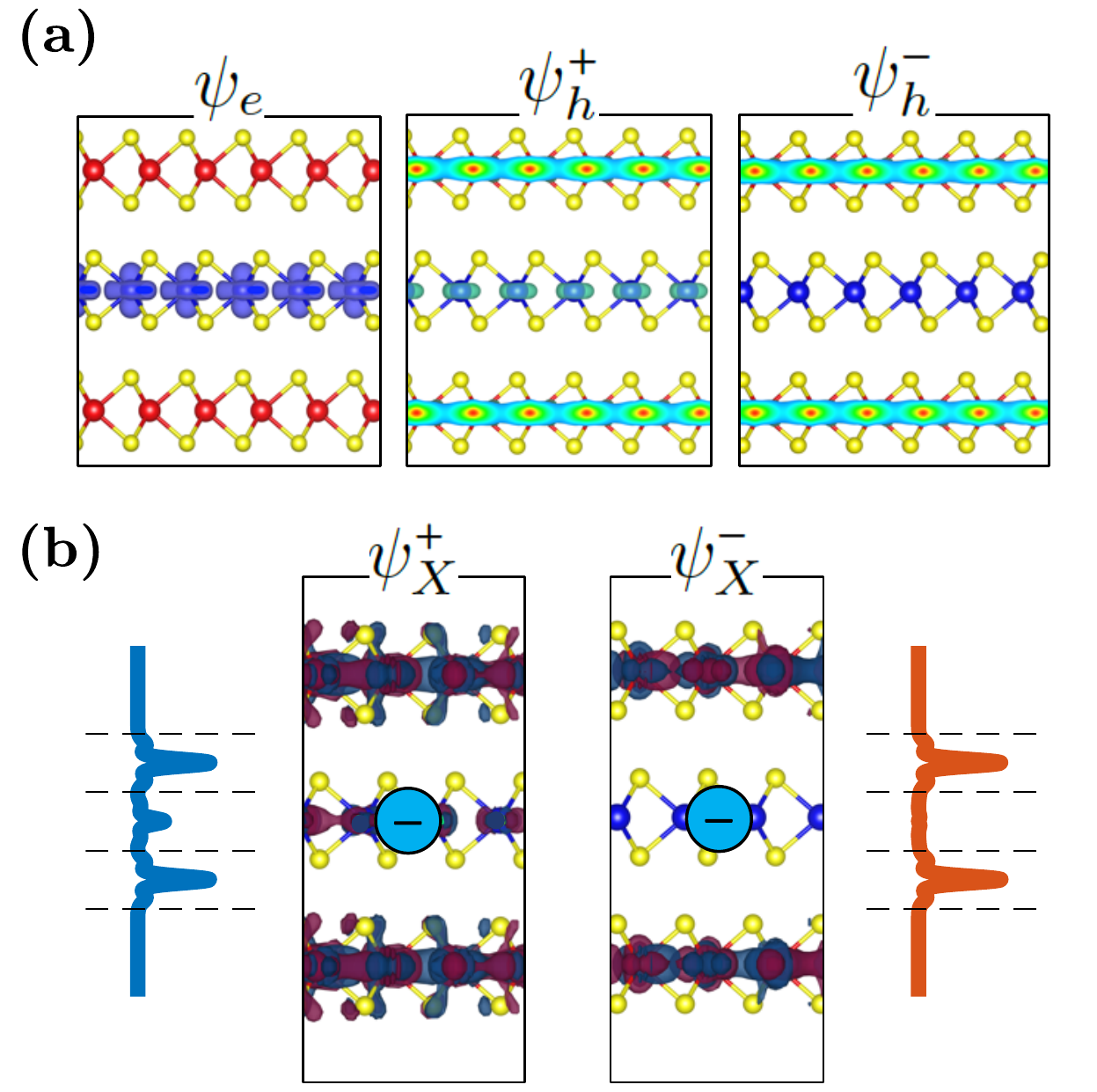}
\caption{\label{fig:gwbse} \textbf{(a)} Quasiparticle wavefunctions of the conduction band ($\psi_e$) and the split valence bands ($\psi^+_h,\psi^-_h$) at K, which construct the low lying excitons. \textbf{(b)} Hole wavefunction of the two lowest-energy exciton states discussed in the main text, $\Psi^+_X$ and $\Psi^-_X$, where the electron position is set to the middle of the cell (blue circle). the schematic wavefunctions represent the constructive and destructive nature of the double-well states composition.}
\end{figure}

We solved the Bethe Salpeter equation (BSE) within the Tamm-Dancoff approximation (TDA) as implemented in the BerkeleyGW code \cite{deslippe2012berkeleygw, rohlfing2000electron}. The BSE matrix elements were calculated in two steps. First we compute electron-hole interactions without including spin-orbit coupling (SOC), on a uniform $24\times24\times1$ k-grid and then interpolated to a $120\times120\times1$ fine k-grid using the Clustered Sampling Interpolation (CSI) technique \cite{felipe2017nonuniform}, in which the BSE matrix elements are explicitly calculated for 10 additional k points in each cluster, sampled along the (110) direction. We used 4 occupied and 4 empty bands in the kernel and absorption calculations. We then correct for spin-orbit effects for the presented excitation energies by reducing the no-SOC exciton binding energy from the computed gap with SOC. The quasiparticle and excitonic states are shown in Fig.~\ref{fig:gwbse}.

\section{\label{Sec:Electrostatic} Calculation of the electrostatic energy of exciton lattices}

In this section we elaborate on the numerical calculation of electrostatic interaction energies associated with various exciton structures considered in the main text. As is standard for a two dimensional Bravais crystal, we span the lattice points, $\vec{R}_{l,m}=l\vec{a}_1+m\vec{a}_2$, using an integer ($\{l,m\}\in \mathbb{Z}^2$) linear combinations of the primitive lattice vectors $\vec{a}_1$ and $\vec{a}_2$. 

The electrostatic interaction energy between a pair of interlayer excitons separated by a in-plane distance $r$ sensitively depends on the relative dipole moment configuration and is given by,
\begin{align}
&E_{p,p}\left(r\right)=\frac{e^{2}}{\kappa}\left(\frac{2}{r}-\frac{2}{\sqrt{r^{2}+d^{2}}}\right)\\
&E_{p,-p}\left(r\right)=\frac{e^{2}}{\kappa}\left(\frac{1}{r}+\frac{1}{\sqrt{r^{2}+\left(2d\right)^{2}}}-\frac{2}{\sqrt{r^{2}+d^{2}}}\right)\\
&E_{q,q}\left(r\right)=\frac{e^{2}}{\kappa}\left(\frac{3}{2}\frac{1}{r}+\frac{1}{2}\frac{1}{\sqrt{r^{2}+\left(2d\right)^{2}}}-\frac{2}{\sqrt{r^{2}+d^{2}}}\right)
\end{align}
Here $E_{p,p}$ and $E_{p,-p}$ represent a dipolar pair excitons with parallel and anti-parallel electric dipole moments, respectively. $E_{q,q}$ represents quadrupole-quadrupole repulsion of two quadrupolar excitons. $d$ is the distance between two neighboring layers, $\kappa$ is the effective dielectric constant, and $e$ is the electron charge.

First, we consider the layer-symmetric (sym) configuration of quadrupolar excitons arranged in a two-dimensional triangular lattice structure.  The standard primitive lattice vectors are $\vec{a}^{\text{tri}}_1=a\{1,0\}$ and $\vec{a}^{\text{tri}}_2=a\{1/2,\sqrt{3}/2\}$. As a function of density the lattice constant equals $a=\sqrt{2/\left(\sqrt{3}n\right)}$. With the above definition electrostatic interaction energy per exciton equals the lattice sum,
\begin{equation}
E^{\text{int}}_{\text{sym}}=\sum_{l,m}' E_{q,q}(|\vec{R}^{\text{tri}}_{l,m}|),
\label{eq:en_q_tri}
\end{equation}
where the restricted sum above does not include the (infinite) self-interaction at $\{l,m\}=\{0,0\}$. At low densities, the main contribution to the sum arises from the the quadrupolar interaction which decays as $r^{-5}$. From dimensional analysis the energy must scale with the density as $E^{\text{int}}_{\text{sym}}\sim n^{5/2}$.

Next, we consider the staggered (stg) configuration, comprising a two-dimensional square lattice of dipolar excitons, where the dipole moments belonging to nearest neighbors sites are anti-parallel. Here, $\vec{a}^{\text{sq}}_1=a\{1,0\}$ and $\vec{a}^{\text{sq}}_2=a\{0,1\}$ and the lattice constant is simply given by $a=1/\sqrt{n}$. The electrostatic interaction energy per exciton equals
\begin{align}\label{eq:E_stg_int}
    E^{\text{int}}_{\text{stg}}=\sum_{n,m}' \delta_{n+m,\text{odd}} E_{p,-p}(|\vec{R}^{\text{sq}}_{n,m}|)+\delta_{n+m,\text{even}} E_{p,p}(|\vec{R}^{\text{sq}}_{n,m}|)
\end{align}

At low densities, the main contribution to the sum is negative and arises from the 
dipolar interaction which decays as $r^{-3}$. As before, from dimensional analysis the energy must scale with the density as $E^{\text{int}}_{\text{sym}}\sim -n^{3/2}$

A numerical calculation of $E^{\text{int}}_{\text{sym}}$ and $E^{\text{int}}_{\text{stg}}$ is presented in fig. 2(a) of the main text.

\section{\label{Sec:Mean-field} Uniform exciton gas}

In this section we consider the mean field approximation and assume no correlation between excitons. In other words we assume all excitons are distributed randomly and independently of each other with an average density $n$. This condition leads to a description of the excitons as a classical gas. Such a description results in higher energy values compared to the two dimensional lattice calculation presented in the main text (fig. 2(a)) and in the previous Section. 

The average number of excitons in an area element $d^{2}r$ is $nd^{2}r$ and the electrostatic interaction energy is: 
\begin{align*} 
&E^{\text{int}}_{\text{stg,gas}} 
                        =\int          d^{2}r\;\left(\frac{n}{2}E_{p,p}\left(r\right)+\frac{n}{2}E_{p,-p}\left(r\right)\right)
                        =\frac{2\pi e^{2}d}{\kappa}n \\
&E^{\text{int}}_{\text{sym,gas}}=     \int d^{2}r\;nE_{q,q}\left(r\right) = \frac{2\pi                                   e^{2}d}{\kappa}n
\end{align*} where $E_{p,p}$ and $E_{p,-p}$ represent the electrostatic interaction two dipolar excitons with parallel and anti-parallel dipole moments, respectively, and $E_{q,q}$ represents quadrupole-quadrupole repulsion of two quadrupolar excitons.

In this approximation, with no spatial correlations between excitons, we find that $E^{\text{int}}_{\text{stg,gas}}=E^{\text{int}}_{\text{sym,gas}}$, and as expected this energy is always positive and larger than the electrostatic energies in the lattice configuration.  (see the dashed black line in fig. 2(a)).

\section{\label{Sec:Square} Square lattice of quadrupolar excitons}

In this section we compare the electrostatic interaction energies of a square versus a triangular lattice of quadrupolar excitons (denoted by $E^{\text{int}}_{\text{sym,sq.}}$ and $E^{\text{int}}_{\text{sym,tr.}}$, respectively).
The expression for the square lattice case is obtained by replacing $\vec{R}^{\text{tri}}_{l,m}\to \vec{R}^{\text{sq}}_{l,m}$ in Eq.~\eqref{eq:en_q_tri}.

Fig.~\ref{fig:triangular_vs_square} depicts a numerical calculation of $E^{\text{int}}_{\text{sym,tr.}}\left(n\right)$ and $E^{\text{int}}_{\text{sym,sq.}}\left(n\right)$. For any given $n$, the electrostatic interaction is minimized for a triangular lattice. Therefore the ground-state lattice symmetry would be triangular in case of a 2D lattice of quatrupolar excitons. 

\renewcommand{\thefigure}{S2}
\begin{figure}
\includegraphics[width=0.45\textwidth]{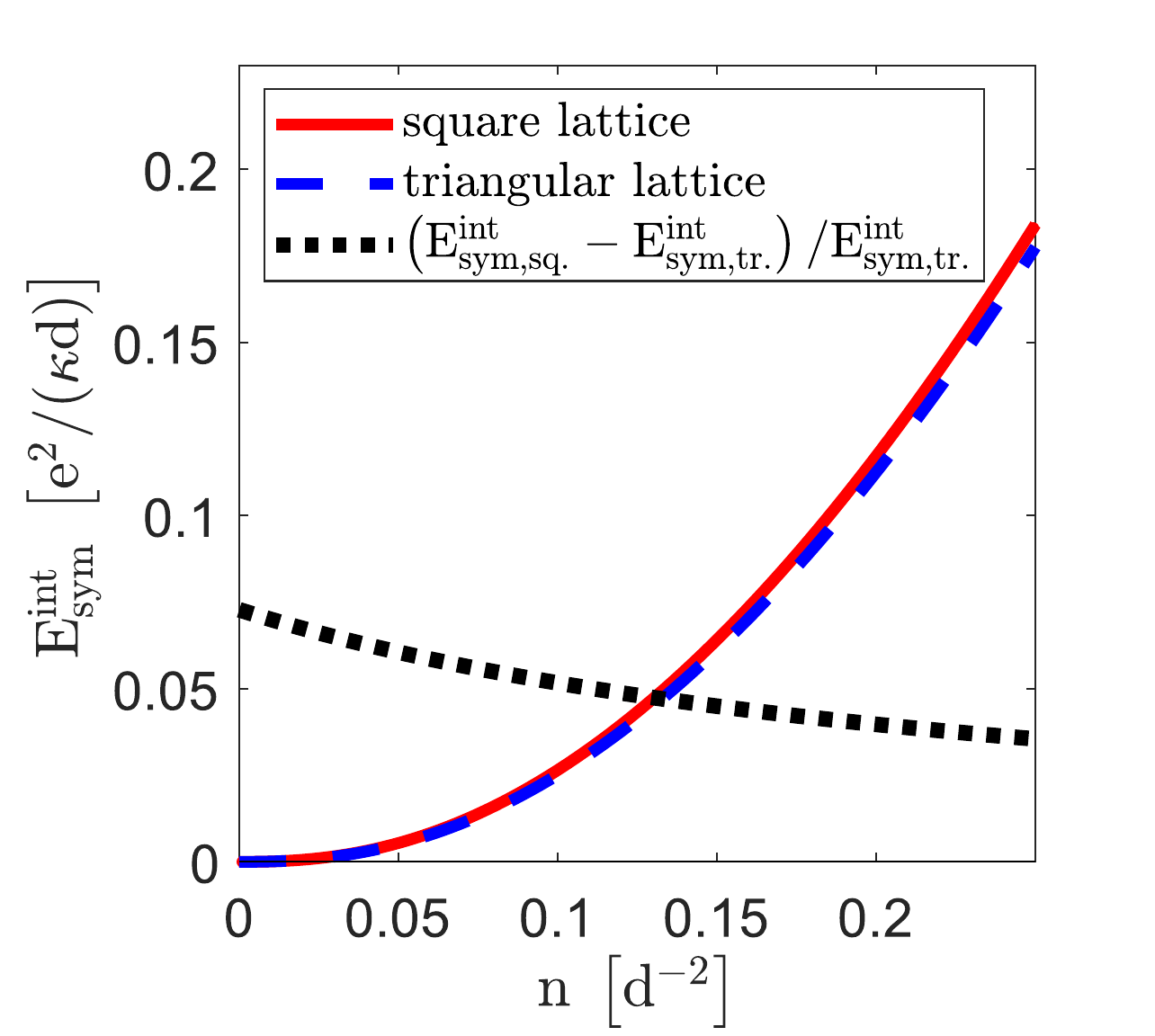}
\caption{\label{fig:triangular_vs_square}  Comparison
between the electrostatic interaction energies of an exciton in a square lattice (dashed blue) and a triangular lattice (solid red) of quadrupolar excitons. The interaction energy in a square lattice is higher by 4-7\% for the presented range of densities (dotted black).}
\end{figure}

\section{Construction of the $T=0$ Phase diagram}

The zero-temperature ground state
of the many particle system is determined in a two-
parameter phase diagram controlled by $R$ and $n$.
For each pair $\left(n,R\right)$ the phase (P.) was determined by the following scheme:
\\

\(\text{P.}\left(n<n_d,R\right) = \begin{cases}
\text{stg droplet,}&E_{\text{sym}}^t\left(n\right)>E_{\text{stg}}^t\left(n_d,R\right)\\
\text{sym lattice,}&E_{\text{sym}}^t\left(n\right)<E_{\text{stg}}^t\left(n_d,R\right)
\end{cases} \)

\(\text{P.}\left(n>n_d,R\right) = \begin{cases}
\text{stg lattice,}&E_{\text{sym}}^t\left(n\right)>E_{\text{stg}}^t\left(n,R\right)\\
\text{sym lattice,}&E_{\text{sym}}^t\left(n\right)<E_{\text{stg}}^t\left(n,R\right)
\end{cases} \)
\\

Schematic illustration of the competing phases is shown if Fig.~\ref{fig:Phases_schematic}.

\renewcommand{\thefigure}{S3}
\begin{figure}
\includegraphics[width=0.48\textwidth]{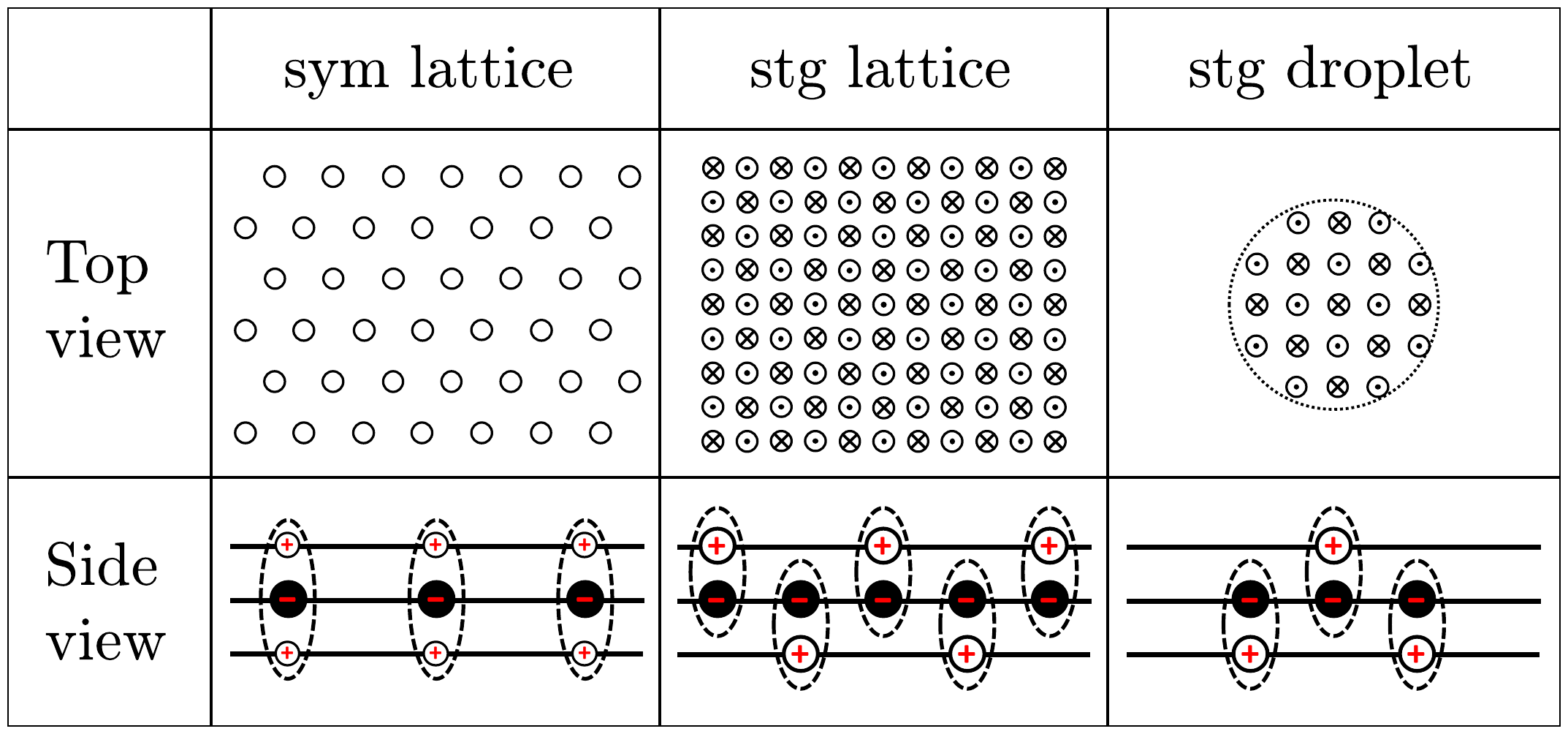}
\caption{\label{fig:Phases_schematic}  Schematic illustration of the competing phases in a three-layer structure with dipolar and quadrupolar excitons.}
\end{figure}

\section{Stability of staggered dipolar droplets}

Dipolar interactions are extended in space and hence one might worry that they may destabilize the formation of droplets. However, as we show below, in the staggered configuration the interaction energy is dominated by only a few adjacent lattice sites.  
To make this statement more quantitative, we plot in Fig.~\ref{fig:stg_NN} a comparison between the full sum $E_{\text{stg}}^{\text{int}}$ from Sec.~\ref{Sec:Electrostatic} (Eq.~\ref{eq:E_stg_int}), and the sum for just nearest and next nearest neighbors, $E_{\text{stg,NN}}^{\text{int}}$. As can be clearly seen, the electrostatic interaction energy of a dipolar exciton in a staggered droplet  $(n<n_d\approx0.12d^{-2})$ is mostly determined by the interaction between nearest and next nearest neighbors, with over 95\% contribution to the total energy. This implies that the energy per particle for an exciton inside a staggered droplet is rather insensitive to the size of the droplet. Consequently, the stability of the droplet with respect to variations in its size is predominantly from excitons at the circumference of a droplet. 

\renewcommand{\thefigure}{S4}
\begin{figure}[h!]
\includegraphics[width=0.45\textwidth]{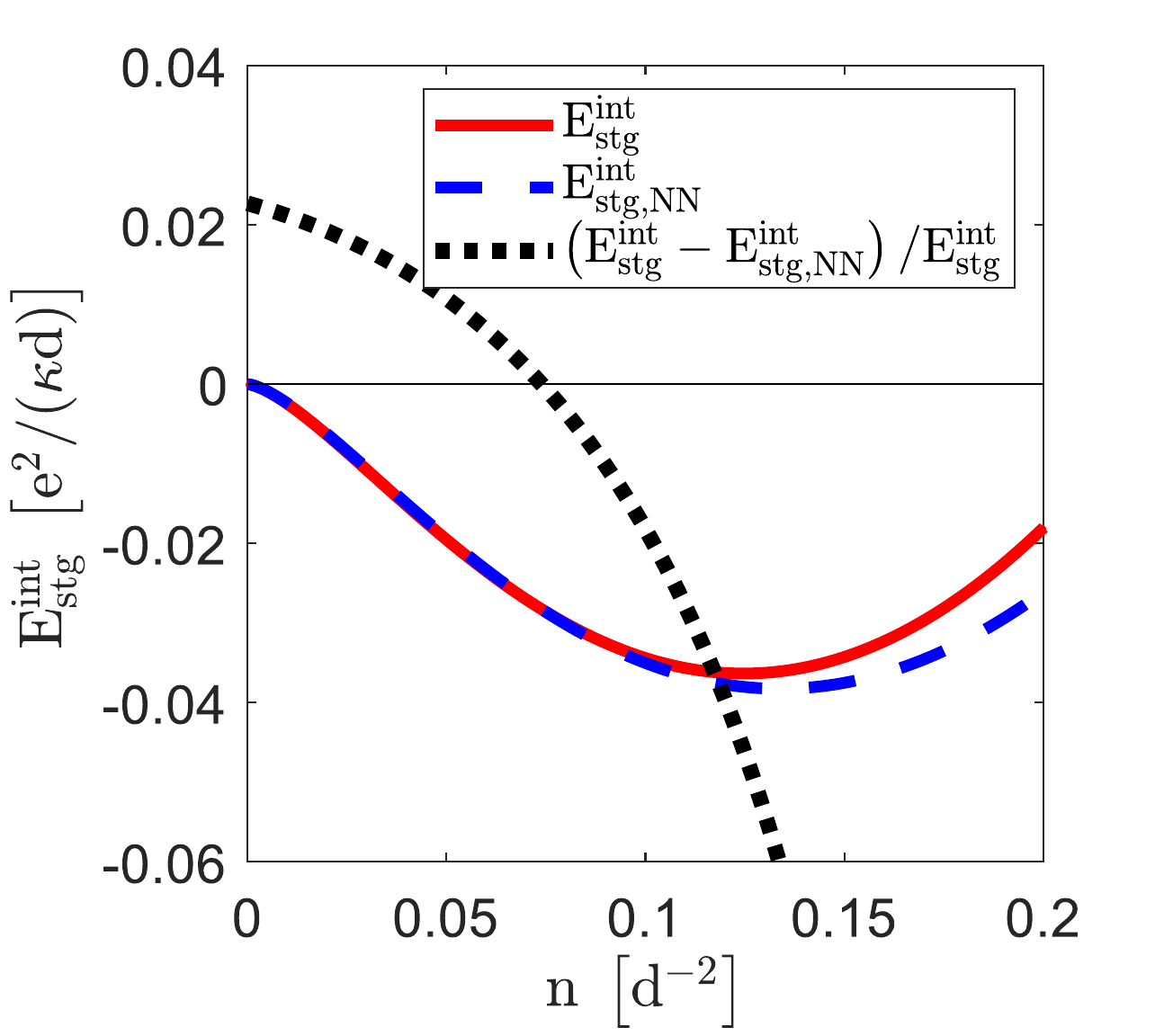}
\caption{\label{fig:stg_NN}  Comparison
between the electrostatic interaction energy of an exciton in a staggered configuration with all other excitons ($E_{\text{stg}}^{\text{int}}$, solid red) and the electrostatic interaction energy with only nearest and next-nearest neighbors ($E_{\text{stg,NN}}^{\text{int}}$, dashed blue). The energy difference is below 5\% for all densities below $n_d$ (dotted black).}
\end{figure}

In a staggered droplet configuration ($n<n_d$), the interaction between neighbors is attractive, and since  excitons on the circumference have less neighbors, excitons on the circumference would have a higher interaction energy with respect to excitons inside the droplet. 
Therefore, the energy of the entire droplet would minimize for a minimum circumference, for a fixed number of excitons at a specific density $n$. As the area of the droplet is conserved,  a circular shape would  minimize the circumference of the droplet. Furthermore, the circumference of a single droplet is smaller than  multiple and smaller in size droplets with the same total area, which means the system would prefer a stable single circular droplet phase, rather then splitting to smaller droplets, which would indicate a possible instability.

\newpage
\bibliography{supplemental_materials}